# Emergence of hadrons from color charge in QCD

**W. K. Brooks[1], H. Hakobyan**
*Universidad Técnica Federico Santa María*
*Avenida España 1680, Valparaíso, Chile*
E-mail: `william.brooks@usm.cl, hayk.hakobyan@usm.cl`

**M. Arratia**
*University of Cambridge*
*Cambridge, England, UK*
E-mail: `miguel.arratia@cern.ch`

**C. Peña**
*California Institute of Technology*
*Pasadena, California, USA*
E-mail: `cristian.morgoth@gmail.com`

The propagation of colored quarks through strongly interacting systems, and their subsequent evolution into color-singlet hadrons, are phenomena that showcase unique facets of Quantum Chromodynamics (QCD). Medium-stimulated gluon bremsstrahlung, a fundamental QCD process, induces broadening of the transverse momentum of the parton, and creates partonic energy loss manifesting itself in experimental observables that are accessible in high energy interactions in hot and cold systems. The formation of hadrons, which is the dynamical enforcement of the QCD confinement principle, is very poorly understood on the basis of fundamental theory, although detailed models such as the Lund string model or cluster hadronization models can generally be tuned to capture the main features of hadronic final states. With the advent of the technical capability to study hadronic final states from lepton scattering with good particle identification and at high luminosity, a new opportunity has appeared. Study of the characteristics of parton propagation and hadron formation as they unfold within atomic nuclei are now being used to understand the coherence and spatial features of these processes and to refine new experimental tools that will be used in future experiments. Fixed-target data on nuclei with lepton and hadron beams, and collider experiments involving nuclei, all make essential contact with these topics and they elucidate different aspects of these same themes. In this paper, a survey of the most relevant recent data and its potential interpretation will be followed by descriptions of planned experiments at Jefferson Lab following the completion of the 12 GeV upgrade, and feasible measurements at a future Electron-Ion Collider.

---

[1] Speaker

## 1. Introduction

The spectacular successes of Quantum Chromodynamics (QCD) in describing the strong interaction at small distance scales are well known. The demonstration of asymptotic freedom [1][2] and the consequent power of perturbative QCD (pQCD) to describe hadronic processes [3] has provided the mandatory foundation for modern searches for the Higgs boson and new physics at hadron colliders such as the LHC [4], where QCD processes need to be very well understood in order to attain the required experimental sensitivity.

A new frontier in QCD is the quantitative understanding of the non-perturbative sector (npQCD). Progress in npQCD, which is the very foundation of nuclear physics, has accelerated enormously over the last decade in the areas of hadron structure and in the study of the thermodynamics of the strong interaction. This progress has taken many forms: a flood of new data from facilities that take advantage of cutting edge accelerator and detector technology [5][6][7][8], advances in theory through lattice QCD from increased computing power, technical breakthroughs, and conceptual breakthroughs [9], as well as advances in related areas such as chiral effective field theory in connecting QCD to nuclei [10][11][12] and other hadrons. These striking advances offer the promise of a groundbreaking revision of our understanding of npQCD over the next two decades.

Due to the confluence of several factors, a new area of opportunity for advancement in npQCD has recently opened up. This opportunity is the quantitative study of the hadronization or fragmentation process in QCD, the topic of this paper. The possibility of performing semi-inclusive deep inelastic scattering on atomic nuclei at high luminosity and with identified final-state hadrons began with the HERMES collaboration [13] in the late 1990's. Prior to that time, experiments with semi-inclusive leptoproduction of undifferentiated hadrons from nuclei were carried out with neutrinos at FNAL, CERN and Serpukhov; with electrons at SLAC; and at CERN and FNAL with high-energy muons by EMC and by E665 respectively [14]. These measurements on nuclear targets had already indicated that hard processes in nuclei were not a simple superposition of scattering from protons and neutrons. Progress beyond that level was inhibited by the lack of identification of the final state hadrons produced following the hard interactions, a technical development that emerged with the operation of the HERMES RICH detector which could identify pions, kaons, and protons up to high momentum. This capability, in combination with nuclear targets and the absence of initial-state interactions offered by the electromagnetic probe, began an exciting new chapter in the quantitative understanding of how hadrons emerge from the color charge of QCD.

Subsequent to the HERMES studies, similar types of data at lower energies, with larger nuclei, and with two orders of magnitude more integrated luminosity have been taken with CLAS at Jefferson Lab; examples of these data will be provided and discussed in the following. Studies in the near future at the upgraded CLAS12 [15] will offer a wider kinematic range, more types of identified hadrons, and another order of magnitude increase in luminosity. The ultimate opportunity to explore the higher-energy themes of hadronization will come at the planned Electron-Ion Collider [16], where a very wide range of kinematics will be available.

## 2. Hadronization in vacuum

The terms hadronization or fragmentation as used here include all aspects of the process by which one or more quarks are removed from a color singlet hadron system via a hard interaction, and subsequently evolve into a system of one or more new color singlet hadrons created out of the energy of the initial interaction. This process is subject to the confinement principle of QCD that demands that all color charge is asymptotically resolved into color singlet subsystems. There are several phenomenological models, such as the Lund string model [17] and the cluster model embedded in Herwig [18], that can describe this process using particular assumptions and constraints from data. However, the constraints from data, such as production rates for different types of particles, have a limited effectiveness in explaining the detailed dynamics of the process, and a solution to the QCD Lagrangian for hadronization is technically far out of reach (although new approaches in lattice QCD may eventually make some headway [9]).

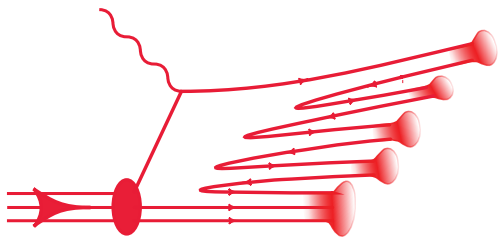

Figure 1. A conceptual sketch of the hadronization process in which a virtual photon interacts with one quark from a nucleon and quark-antiquark pairs emerging from the vacuum subsequently join to form hadrons.

The process is visualized as, for example in Deep Inelastic Scattering (DIS), a colored struck quark separating from the remnants of a nucleon while still feeling the constant force due to the pull of a flux tube or color string connecting the struck quark to the remnant system. In this context, quark-antiquark pairs emerge from the vacuum and evolve into color singlet hadrons, a natural picture for forming mesons but less natural for baryons. Figure 1 shows a schematic conceptual sketch of this process, which is fundamentally non-perturbative. In a perturbatively-inspired view of the process, the struck quark erratically emits gluons at a high rate, losing energy in vacuum at a rate comparable to the string constant of the Lund string model, ~1 GeV/fm, and forming a longitudinally-extended pool of color charge out of which color singlet hadrons can condense.

In modern models of these processes, a list of final state quarks and gluons in the QCD cascade is made from either a parton shower calculation, or a pQCD matrix element calculation, and then in a separate step those quarks and gluons form into hadrons via, for example, the Lund string model. This concept of first producing all partons and later producing hadrons is an artificial separation, however, for a vacuum process there is no way to obtain more information to constrain the description further.

The process of encapsulating the experimental cross section into universal parton distribution functions and universal fragmentation functions is a highly developed science that has its basis in the QCD factorization theorems [19]. Perhaps because the process can be characterized in this way, without knowing many of the details of the dynamics, and because available data did not allow further insights into these dynamics, little attention has been paid in recent years to pushing the envelope of our knowledge about hadronization. Now, with the advent of high-luminosity leptoproduction experiments with nuclei and identified hadrons, it is possible to make new progress into understanding this mysterious, fascinating, and ubiquitous process that is so intimately tied to the confinement principle of QCD.

## 3. Hadronization in medium

Parton propagation and hadronization *in-medium* offer the potential to provide new information about the details of the propagation and hadronization process through the interactions of the hadronization constituents with the medium. In the hot systems formed in heavy ion collisions, the phenomenon of jet quenching is described as arising from parton energy loss primarily via gluon bremsstrahlung in-medium, in addition to collisional losses that are particularly important for heavy quarks. In cold nuclei, the radiative energy loss is smaller and there may also be a finite probability that the color-singlet system forms within the nucleus. In the current descriptions of the HERMES data as well as historical work [20], these two processes have been invoked to describe the observed suppression of hadron production: 1) energy loss of the struck quark by gluon emission [21][22][23], and 2) inelastic interaction of the (pre-) hadron with the nucleons in the nucleus [24][25]. Since the hadronic cross section is large (with a mean free path of order 1 fm) even a hadron in formation could in principle interact appreciably, albeit potentially with a reduced cross section. Thus, one speaks of "pre-hadrons" interacting as well as fully-formed hadrons.

There are many new experimental tools which can be employed to understand these kinds of data. In the following sections of this paper, only three of these are developed, however, it is worth mentioning a longer list here.

(a) The modification of the parton propagation and hadronization processes can be quantified by the use of the hadronic multiplicity ratio $R_h$:

$$R_h = \frac{\frac{1}{N_e^A} N_h^A}{\frac{1}{N_e^D} N_h^D}$$

The numerator of this ratio is given by the number of hadrons of type h produced from nucleus A, differential in up to 4 variables[2] such as $Q^2$, $\nu$, z, and $p_T$, normalized to the number of scattered electrons. The denominator is the same expression for a small nucleus, in this example, deuterium. If no nuclear effects are present (including the EMC effect) the ratio is equal to unity.

(b) The broadening of the distribution of hadron momentum transverse to the direction of the virtual photon, $p_T$, carries unique information on the interaction of the propagating parton with the gluon field of the nucleus, $G_A(x,Q^2)$. It is defined as the difference between the mean value of the distribution of the square of $p_T$ of the heavy nucleus with that of the light nucleus:

$$\Delta p_T^2 \equiv \langle p_T^2 \rangle_A - \langle p_T^2 \rangle_D$$

(c) A direct measurement of energy loss may be possible by looking at the distributions in energies of the produced hadrons in selected kinematic conditions.

The above three observables are discussed in more detail below. Additional observables that will provide closely related information on these physics topics in the future include:

(d) Two-hadron attenuation may clarify hadronization mechanisms [26].
(e) Photon-hadron correlations may reveal a new medium transport coefficient [23].
(f) Bose-Einstein correlations access spatial and temporal properties of hadronization.
(g) Observation of low-energy associated particles produced in the medium [27].
(h) The target fragmentation region may be studied via interactions in nuclei.

---

[2] Traditionally, the fifth variable $\phi_{pq}$ is integrated over when defining leptoproduction multiplicity.

(i) Heavy quark energy loss via heavy meson production at present/future facilities [16].
(j) Single and double spin asymmetries in meson production from nuclei [28].
(k) Color transparency accessed through, e.g., nuclear vector meson production [29].

## 4. Recent experimental results

In Fig. 2 are shown preliminary three-fold differential multiplicity ratios for positive pion production in CLAS on targets of carbon, iron, and lead relative to deuterium as a function of $z_\pi = E_\pi/\nu$ with $E_\pi$ the pion energy. This observable is defined in (a) above. The pattern seen is similar to the one seen at HERMES, here extended to heavier targets at lower energies. The characteristic pattern of increasing suppression for heavier nuclei is seen, indicating a path length dependent process, and the rise above 1.0 at low z suggests a contribution to the suppression consisting of a shift to lower z, as would clearly be expected for gluon radiation and also to some extent could be due to inelastic collisions of the prehadron. For lead, the heaviest nucleus, the suppression is more than 50% for more than half of the z range. The expected suppression at highest z is observed, a consequence of energy conservation (hadrons with z=1 cannot have deposited *any* energy in the medium). A similar pattern has been seen recently for $K^0_s$ [30].

This observable includes, among other things, important information on the time development of the hadronization process. In a comparison of data from a broad range of lepton energies it was found [25] that time dilation of the fundamental processes, in combination with a well-tested and realistic model of hadron interactions in nuclei, was able to explain the dependence of the data on the energy transfer $\nu$, which has the natural interpretation for x>0.1 as the initial quark energy.

Besides being amenable to theoretical description in models that treat parton energy loss and prehadron interactions, these data are also relevant to a broad variety of topics, such as attenuation of low-momentum hadrons in proton-lead scattering at the LHC or in deuteron-gold scattering at RHIC, and in understanding and constraining neutrino-nucleus deep inelastic scattering. Further, observables of this type will lead to important new information when used at the future Electron Ion Collider, particularly in comparisons of heavy meson production. Due to the unique features of heavy quark fragmentation functions, new and important information will be accessible on heavy quark energy loss in cold matter.

Because the energies are low for these data, it is important to clarify which features of the data correspond to partonic processes and which pertain to hadronic processes, when possible. The $p_T$ broadening observable, defined in (b) above, offers some unique information on this point. In the left-hand panel of Fig. 3 is shown an example of the $p_T$ broadening observable for positive pions plotted as a function of the energy of the outgoing pion for the same three nuclei. In the right-hand panel is a parameterization of the well-known pion-nucleon cross section weighted according to the neutron-to-proton composition of carbon (top red line), iron, and lead (bottom black line). In the case that the origin of the $p_T$ broadening is *hadronic* in nature, one would expect that the $p_T$ broadening would exhibit an enhancement at the same energy, and that the shape of the distribution in hadron energy would reflect the shape of the $\pi$-N cross section. However, no hint of an enhancement is seen, and the rising of the broadening up to the pion energy of 1.5 GeV is in contradiction with the constant or falling $\pi$-N cross section. Thus we are driven to the tentative and perhaps surprising conclusion that the broadening process is a fully partonic one, and that the formation of the hadron occurs over a distance considerably longer than nuclear dimensions, or that the cross section of the forming hadron with the medium is very strongly suppressed. While this may not have been anticipated in advance of the data, it is good news since it means that $p_T$ broadening can be used as a tool to probe partonic aspects of hadronization even at quite low energies.

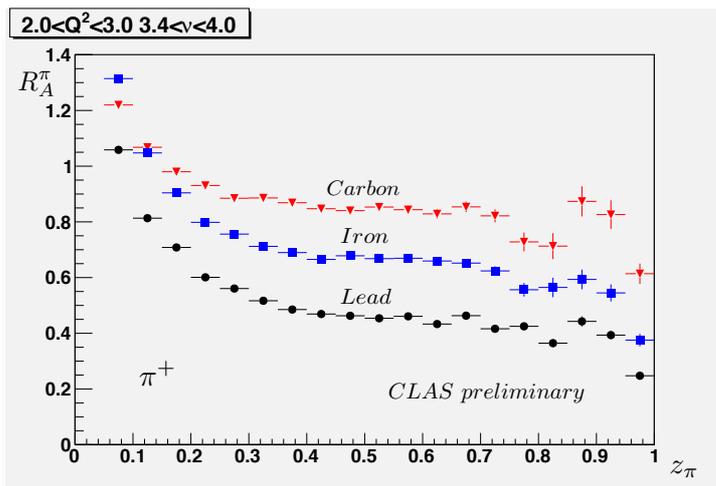

Figure 2. Preliminary threefold differential multiplicity ratio data from CLAS for heavier targets relative to deuterium. The figure indicates the dependence of this ratio on the value of z for the pion in a bin in momentum transfer $Q^2$ and in energy transfer $\nu$. Inverted triangles indicate carbon, squares indicate iron, and filled circles indicate lead. Error bars indicate statistical uncertainties only.

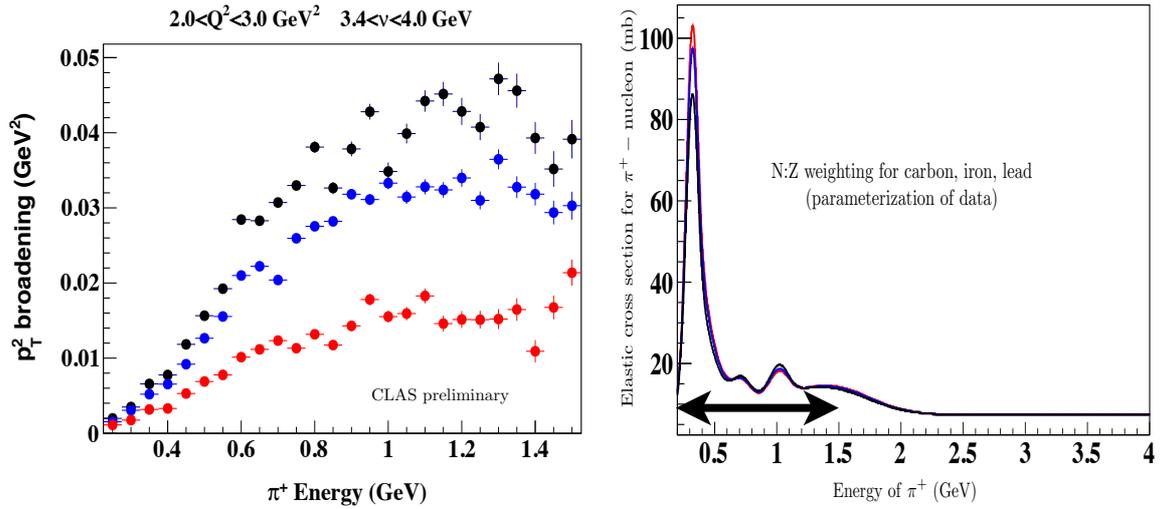

Figure 3. In the left panel is depicted the pattern of transverse momentum broadening of the pion as a function of pion energy for carbon (bottom points), iron, and lead (top points). In the right panel is shown is a parameterization of the pion-nucleon elastic scattering cross section for (top to bottom) carbon, iron, and lead; the arrow on the bottom shows the energy range of this plot that is also shown on the left panel. The pronounced peak seen in the right panel, as well as the enhancements up to 2 GeV in pion energy, would be expected to be visible in the left panel spectrum if the broadening arises from scattering of pions within the nuclear medium. Since no hint of the large peak is visible in the left panel spectrum, it is argued that the broadening is essentially purely partonic, and that the formation of the hadron in these data takes place over a distance that exceeds nuclear dimensions.

The final example of a study that can shed light on parton propagation in the medium is the direct measurement of parton energy loss, mentioned in (c) above. This method has never been used previously, and thus this is an exploratory effort, however, current results seem promising. Figure 4 shows the pion energy spectra for deuterium superimposed on the pion energy spectrum from lead, which has been shifted horizontally along the axis to simulate an average energy loss for the pions caused by energy loss by the struck quark. The data are normalized to unity in order to compare the spectral shapes. A requirement that Feynman x ($x_F$) is greater than 0.1 is applied to emphasize the current fragmentation region; this requirement affects the shape of the spectrum on the lower-energy edge. The value of $x_F$ for the heavy nucleus is calculated by shifting the pion energy in the equation for $x_F$ by the putative energy loss in the heavy nucleus, and the method is iterated for varying values of the energy loss shifts.

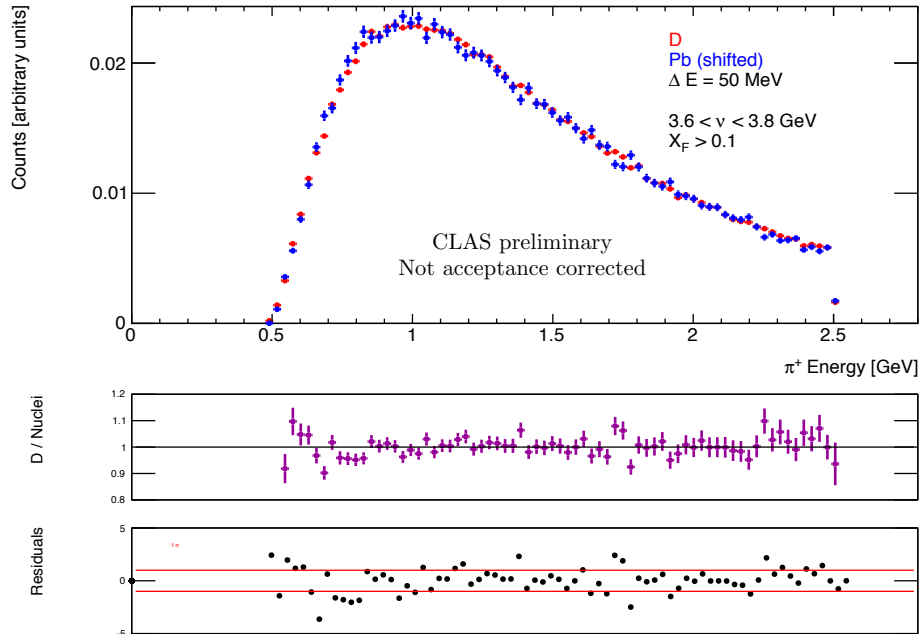

Figure 4. The positive pion energy spectrum for deuterium overlaid with the positive pion energy spectrum from lead shifted along the horizontal axis; normalized to unit area. These data are plotted for a narrow bin in energy transfer and for Feynman x greater than 0.1; the acceptance corrections, while expected to be small, have not been applied. The shapes of the two spectra are seen to be virtually identical within the uncertainties. A Kolmogorov test indicates the shapes are the same at greater than a 95% confidence level.

An optimum shift for these data, which still lack acceptance corrections and radiative corrections, is found which gives a shape match that indicates the two spectra are the same. For every shift a Kolmogorov compatibility test is performed between the two distributions (deuterium and shifted heavy nuclei); the values of energy shift that make the two distributions statistically compatible at >95% confidence level are taken as possible energy loss values. This study permits to reject all other possible energy loss values, resulting in upper and lower bounds of preferred values. An increase of energy loss is found with increasing nuclear size, as expected.

The tentative interpretation of these data is that the shift in pion energy is due to a shift in the struck quark energy caused by gluon radiation in the medium and averaged over the nuclear path lengths. The identicality of the shapes from the two spectra suggests that the fragmentation process takes place primarily outside of the nucleus, or else that it otherwise is not significantly affected by the medium. This result echos observations with jets at the LHC in heavy ion collisions [31] where ratios of jet fragmentation functions for different centralities are plotted over the full range in z. Figure 4, with its narror range of $\nu \approx 3.7$ GeV, is equivalent to a plot of z with data from z = 0.13 to z = 0.67 with the $x_F$ cut applied. If one takes the *ratio* of the unshifted Pb spectrum to the deuterium spectrum, the ratio will be below 1.0 for 0.25<z<0.67 and above 1.0 for lower z, as a simple consequence of the spectral shapes. This is qualitatively the same pattern seen at LHC for jets in Pb+Pb collisions.

The value of the pion energy shift is expected to be smaller than the average quark energy loss, nominally by a factor of $<z^{-1}> \approx 3$, and in principle the shape of the fragmentation function can also be affected by the quark energy loss. However, for the small shifts observed it can be argued this is a tiny effect that can easily be modeled.

The early results from this exploratory effort suggest much smaller energy losses than would be given by simple expressions found in the literature. However, it should be remembered that the typical focus of attention in the literature is on the high energy limits of the pQCD theory, and often as applied to hot dense systems [32][33]. Indeed, for systems of fixed size L, the total energy loss is approximately constant in energy above a critical energy $E_{crit}$, and reduces below that value, as is schematically indicated in Fig. 5 [34]. Taking the value of $E_{crit}$ as the square of the system size times a transport coefficient of 0.08 GeV$^2$/fm, and using 2.5 fm and 6.5 fm as the typical system length for carbon and lead respectively, one arrives at a critical energy range of 2.5 GeV to 17 GeV, ignoring any additional coefficients for $E_{crit}$. Thus, the CLAS data with $\nu \sim$2-4 GeV on carbon, aluminum, iron, and lead are likely to fall on both sides of the critical energy: below for the two heavy targets, and above for the two light targets. If the above direct measurement of quark energy loss is ultimately proven to be reliable, a full experimental study of the curve in Fig. 5 is feasible with the upgraded CLAS12 in the next few years, profoundly increasing the experimental understanding of partonic energy loss in QCD.

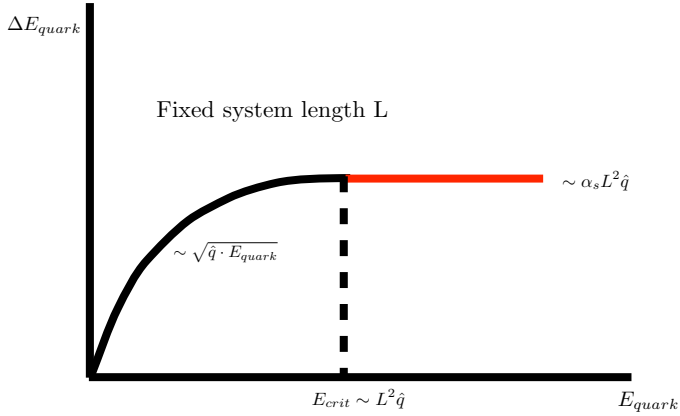

Figure 5. Behavior of quark energy loss for a medium of fixed length L. Above the critical energy, the total energy loss is quadratic in system length and constant in energy, but below the critical energy the total energy loss decreases as the square root of the quark energy. Because the critical energy itself depends on the path length, the average value of the critical energy is smaller for light nuclei and larger for heavy nuclei.

## 5. Conclusions

The propagation of colored quarks through strongly interacting systems, and their subsequent evolution into color-singlet hadrons, are phenomena that showcase unique facets of Quantum Chromodynamics (QCD). With the advent of the technical capabilities to study hadronic final states from lepton scattering with good particle identification and high luminosity, a new opportunity has appeared to significantly impact our understanding of npQCD, the foundation of nuclear physics, in the area of hadronization physics. Study of the characteristics of parton propagation and hadron formation as they unfold within atomic nuclei are now being used to understand the coherence and spatial features of hadronization and to refine new experimental tools that will be used in future experiments. Fixed-target data on nuclei with lepton and hadron beams, and collider experiments involving nuclei, all make essential contact with these topics and they elucidate different aspects of these same themes. In this paper, samples of the most relevant recent data and its potential interpretation were discussed. A suppression of charged pions at the level of more than 50% was observed for the lead nucleus over a wide range in the z variable, and the systematic behavior is qualitatively consistent with previous measurements at HERMES. Studies with $p_T$ broadening were shown to suggest that this observable successfully isolates partonic behavior, and that the formation of the hadron is apparently longer than nuclear dimensions for these kinematics. Exploratory studies of direct measurements of quark energy loss appear promising, and suggest that the fragmentation process takes place primarily outside of the nucleus, or else that it otherwise is not significantly affected by the medium, an observation that echos results for jets at the LHC in heavy ion collisions. Further, the early results of these studies suggest rather small energy losses, which is qualitatively consistent with pQCD due to the lower parton energies involved. Besides being amenable to theoretical description in models that treat parton energy loss and prehadron interactions, these data are also relevant to a broad variety of topics, such as attenuation of low-momentum hadrons in proton-lead scattering at the LHC, and in understanding and constraining neutrino-nucleus deep inelastic scattering. Substantial extensions of these studies will be carried out first with CLAS12 following the completion of the 12 GeV upgrade, and later with a much broader range of kinematics at a future Electron-Ion Collider.

## 6. Acknowledgements

The authors would like to thank S. Peigne for useful discussions concerning partonic energy loss in pQCD, and F. Arleo for useful discussions about implementing energy loss in theoretical descriptions of data. We gratefully acknowledge support from the following Chilean grants: BASAL FB0821, CONICYT ACT-119, and FONDECYT #1120953 and #11121448.